\documentclass{ws-procs10x7}

\usepackage{epsfig}

\begin{document}

\title{Dark energy and dark matter from cosmological observations}

\author{Steen Hannestad}

\address{Department of Physics and Astronomy, University of Aarhus, Ny Munkegade,
DK-8000 Aarhus C, Denmark, E-mail: sth@phys.au.dk}

\twocolumn[\maketitle\abstract{The present status of our knowledge
about the dark matter and dark energy is reviewed. Bounds on the
content of cold and hot dark matter from cosmological observations
are discussed in some detail. I also review current bounds on the
physical properties of dark energy, mainly its equation of state
and effective speed of sound.}]

\section{Introduction}

The introduction of new observational techniques has in the past
few years moved cosmology into the era of precision science. With
the advent of precision measurements of the cosmic microwave
background (CMB), large scale structure (LSS) of galaxies, and
distant type Ia supernovae, a new paradigm of cosmology has been
established. In this new standard model, the geometry is flat so
that $\Omega_{\rm total} = 1$, and the total energy density is
made up of matter ($\Omega_m \sim 0.3$) [comprised of baryons
($\Omega_b \sim 0.05$) and cold dark matter ($\Omega_{\rm CDM}
\sim 0.25$)], and dark energy ($\Omega_X \sim 0.7$). With only a
few free parameters this model provides an excellent fit to all
current observations
\cite{Riess:1998cb,Perlmutter:1998np,Spergel:2003cb,sdss1}.
However, cosmology is currently very much a field driven by
experiment, not theory. While all current data can be described by
a relatively small number of fitting parameters the understanding
of the underlying physics is still limited.

Here, I review the present knowledge about the observable
cosmological parameters related to dark matter and dark energy,
and relate them to the possible underlying particle physics
models. I also discuss the new generation of experiments currently
being planned and built, particularly those designed to measure
weak gravitational lensing on large scales. These instruments are
likely to bring answers to at least some of the fundamental
questions about dark matter and dark energy.

\section{Cosmological data}

\subsection{Large Scale Structure (LSS).}

At present there are two large galaxy surveys of comparable size,
the Sloan Digital Sky Survey (SDSS) \cite{sdss1,sdss2} and the
2dFGRS (2~degree Field Galaxy Redshift Survey) \cite{2dFGRS}. Once
the SDSS is completed in December 2005 it will be significantly
larger and more accurate than the 2dFGRS, measuring in total about
$10^6$ galaxies.

Both surveys measure angular positions and distances of galaxies,
producing a fully three dimensional map of the local Universe.
From this map various statistical properties of the large scale
matter distribution can be inferred.

The most commonly used is the power spectrum $P(k,\tau)$, defined
as
\begin{equation}
P(k,\tau) = |\delta_k|^2(\tau),
\end{equation}
where $k$ is the Fourier wave number and $\tau$ is conformal time.
$\delta$ is the $k$'th Fourier mode of the density contrast,
$\delta \rho/\rho$.

The power spectrum can be decomposed into a primordial part,
$P_0(k)$, generated by some mechanism (presumably inflation) in
the early universe, and a transfer function $T(k,\tau)$,
\begin{equation}
P(k,\tau) = P_0(k) T(k,\tau).
\end{equation}
The transfer function at a particular time is found by solving the
Boltzmann equation for $\delta(\tau)$ \cite{MB}.

As long as fluctuations are Gaussian, the power spectrum contains
all statistical information about the galaxy distribution. On
fairly large scales $k \leq 0.1 \, h$/Mpc this is the case, and
for that reason the power spectrum is the form in which the
observational data is normally presented.

\subsection{Cosmic Microwave Background.}

The CMB temperature fluctuations are conveniently described in
terms of the spherical harmonics power spectrum $C_l^{TT} \equiv
\langle |a_{lm}|^2 \rangle$, where $\frac{\Delta T}{T}
(\theta,\phi) = \sum_{lm} a_{lm}Y_{lm}(\theta,\phi)$.  Since
Thomson scattering polarizes light, there are also power spectra
coming from the polarization. The polarization can be divided into
a curl-free ($(E)$) and a curl ($(B)$) component, much in the same
way as $\vec{E}$ and $\vec{B}$ in electrodynamics can be derived
from the gradient of a scalar field and the curl of a vector field
respectively (see for instance \cite{Kamionkowski:1996ks} for a
very detailed treatment). The polarization introduced a sequence
of new power spectra, but because of different parity some of them
are explicitly zero. Altogether there are four independent power
spectra: $C_l^{TT}$, $C_l^{EE}$, $C_l^{BB}$, and the $T$-$E$
cross-correlation $C_l^{TE}$.

The WMAP experiment has reported data only on $C_l^{TT}$ and
$C_l^{TE}$ as described in
Refs.~\cite{Bennett:2003bz,Spergel:2003cb}. Other experiments,
while less precise in the measurement of the temperature
anisotropy and not providing full-sky coverage, are much more
sensitive to small scale anisotropies and to CMB polarization.
Particularly the ground based CBI \cite{Pearson:2002tr}, DASI
\cite{Kovac:2002fg}, and ACBAR \cite{Kuo:2002ua} experiments, as
well as the BOOMERANG balloon experiment
\cite{Jones:2005yb,Piacentini:2005yq,Montroy:2005yx} have provided
useful data.

\subsection{Type Ia supernovae}

Observations of distant supernovae have been carried out on a
large scale for about a decade. In 1998 two different projects
almost simultaneously published measurements of about 50 distant
type Ia supernovae, out to a redshift or about 0.8
\cite{Riess:1998cb,Perlmutter:1998np}. These measurements were
instrumental for the measurement of the late time expansion rate
of the universe.

Since then a, new supernovae have continuously been added to the
sample, with the Riess et al. \cite{Riess:2004} "gold" data set of
157 distant supernovae being the most recent. This includes
several supernovae measured by the Hubble Space Telescope out to a
redshift of 1.7.

\section{Cosmological parameters}

Based on the present cosmological data, many different groups have
performed likelihood analyses based on various versions of the
standard Friedmann-Robertson-Walker cosmology (see for instance
\cite{sdss1,Seljak:2004xh} for recent analyses). A surprisingly
good fit is provided by a simple, geometrically flat universe, in
which 30\% of the energy density is in the form of
non-relativistic matter and 70\% in the form of a new, unknown
dark energy component with strongly negative pressure. Fig.~1
shows the allowed region from a combined fit of WMAP, SDSS, and
Type-Ia supernova data.

\begin{figure}[h]
\begin{center}
\epsfysize=6.5truecm\epsfbox{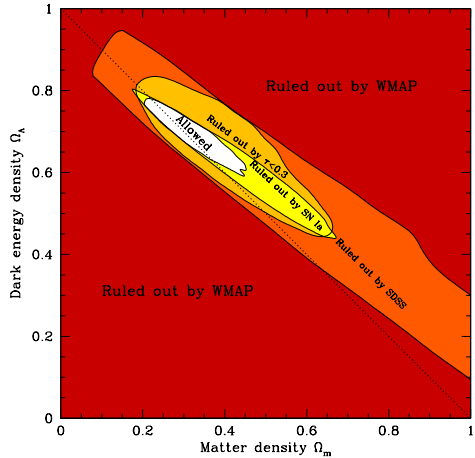} \vspace{0.5truecm}
\end{center}
\caption{The 95\% likelihood contour for $\Omega_m$ and
$\Omega_\Lambda$ from WMAP, SDSS, and SNI-a data [with permission
from \protect\cite{sdss1}]} \label{max}
\end{figure}

In its most basic form, the dark energy is in the form of a
cosmological constant where $w \equiv P/\rho = -1$. The only free
parameters in this model are: $\Omega_m$, the total matter
density, $\Omega_b$, the density in baryons, and $H_0$, the Hubble
parameter. In addition to these there are parameters related to
the spectrum of primordial fluctuations, presumably generated by
inflation. Observations indicate that the fluctuations are
Gaussian and with an almost scale invariant power spectrum. More
generally, the primordial spectrum is usually parameterized by two
parameters: A, the amplitude, and $n_s$ the spectral tilt of the
power spectrum. Finally, there is the parameter $\tau$ which is
related to the redshift of reionization of the Universe.
Altogether, standard cosmology is describable by only 6 parameters
(5 if the spectrum is assumed to be scale invariant \footnote{See
\cite{Liddle:2004nh,Mukherjee:2005wg} for a discussion about how
to estimate the number of cosmological parameters needed to fit
the data.}.

Adding other parameters to the fit does not significantly alter
the determination of the 6 fundamental parameters, although in
some cases the estimated error bars can increase substantially.

\section{Dark matter}

The current cosmological data provides a very precise bound on the
physical dark matter density \cite{sdss1}
\begin{equation}
\Omega_m h^2 = 0.138 \pm 0.012,
\end{equation}
although this bound is somewhat model dependent. It also provides
a very precise measurement of the cosmological density in baryons
\cite{sdss1}
\begin{equation}
\Omega_b h^2 = 0.0230^{+0.0013}_{-0.0012}.
\end{equation}
This value is entirely consistent with the estimate from Big Bang
nucleosynthesis, based on measurements of deuterium in high
redshift absorption systems, $\Omega_b h^2 = 0.020 \pm 0.002$
\cite{Burles:2000zk,Cyburt:2003fe}.

The remaining matter density consists of dark matter with the
density \cite{sdss1}
\begin{equation}
\Omega_{\rm dm} h^2 = 0.115 \pm 0.012.
\end{equation}
The bound on the dark matter density in turn provides strong input
on any particle physics model for dark matter. Space limitations
allow only for a very brief review of the cosmological constraints
on dark matter. Very detailed reviews can be found in
\cite{Bergstrom:2000pn,Bertone:2004pz}.

\subsection{WIMPs}

The simplest model for cold dark matter consists of WIMPs - weakly
interacting massive particles. Generic WIMPs were once in thermal
equilibrium, but decoupled while strongly non-relativistic. For
typical models with TeV scale SUSY breaking where neutralinos are
the LSPs, one finds that $T_D/m \sim 0.05$. SUSY WIMPs are
currently the favoured candidate for cold dark matter (see
\cite{Bertone:2004pz}). The reason is that for massive particles
coupled to the standard model via a coupling which is suppressed
by 1/TeV and with a mass of order 100 GeV to 1 TeV a present
density of $\Omega_m h^2 \sim 0.1$ comes out fairly naturally.
SUSY WIMPs furthermore have the merit of being detectable. One
possibility is that they can be detected directly when they
deposit energy in a detector by elastically scattering (see the
contribution by Laura Baudis to these proceedings). Another is
that WIMPs annihilate and produce high energy photons and
neutrinos which can subsequently be detected (see the contribution
by Rene Ong to these proceedings).

\subsection{CDM Axions}

WIMPs are by no means the only possibility for having cold dark
matter. Another possibility is that CDM is in the form of axions,
in which case the mass needed to produce the correct energy
density is of order $10^{-3}$ eV. In this case the axions would be
produced coherently in a condensate, effectively acting as CDM
even though their mass is very low (see for instance
\cite{Raffelt:2005mt} for a recent overview).

\subsection{Exotica}

Another interesting possibility is that dark matter consists of
very heavy particles. A particle species which was once in thermal
equilibrium cannot possible be the dark matter if its mass is
heavier than about 350 TeV \cite{Griest:1989wd}. The reason is
that its annihilation cross section cannot satisfy the unitarity
bound. Therefore, heavy dark matter would have to be produced out
of thermal equilibrium, typically by non-perturbative processes at
preheating towards the end of inflation (see for instance
\cite{Chung:1998ua}). These models have the problem of being
exceedingly hard to verify or rule out experimentally.

\subsection{Hot dark matter}

In fact the only dark matter particle which is known to exist from
experiment is the neutrino. From measurements of tritium decay,
standard model neutrinos are known to be light. The current upper
bound on the effective electron neutrino mass is 2.3 eV at 95\%
C.L. \cite{mainz} (see also the contribution by Christian
Weinheimer to these proceedings). Such neutrinos decouple from
thermal equilibrium in the early universe while still
relativistic. Subsequently they free-stream until the epoch around
recombination where they become non-relativistic and begin to
cluster. The free-streaming effect erases all neutrino
perturbations on scales smaller than the free-streaming scale. For
this reason neutrinos and other similar, light particles are
generically known as hot dark matter. Models where all dark matter
is hot are ruled out completely by present observations, and in
fact the current data is so precise that an upper bound of order 1
eV can be put on the sum of all light neutrino masses
\cite{numass,numass2,numass3,numass4,numass5,numass6,numass7,numass8,numass9,numass10,sth2}.
This is one of the first examples where cosmology provides a much
stronger constraint on particle physics parameters than direct
measurements. The robustness of the neutrino mass bound has been a
topic many papers over the past two years. While some derived mass
bounds, as low as 0.5 eV are almost certainly too optimistic to
consider robust at present, it is very hard to relax the upper
bound to much more than 1.5 eV \cite{sth2}. The reason for the
difference in estimated precision lies both in the assumptions
about cosmological parameters, and in the data sets used.

In the future, a much more stringent constraint will be possible,
especially using data from weak lensing (see section 6).

\subsection{General thermal relics}

The arguments pertaining to neutrinos can be carried over to any
thermal relic which decoupled while relativistic. As long as the
mass is in the eV regime or lower the free streaming scale is
large than the smallest scales in the linear regime probed by LSS
surveys. This has for instance been used for particles such as
axions \cite{mirizzi}. It should of course be noted that these
axions are in a completely different different mass range than the
axions which could make up the CDM. At such high masses, the
axions would be in thermal equilibrium in the early universe until
after the QCD phase transition at $T \sim 100$ MeV and therefore
behave very similarly to neutrinos.

However, for relics which decouples very early, the mass can be in
the keV regime. In that case it is possible to derive mass bounds
using data from the Lyman-$\alpha$ forest which is at much higher
redshifts and therefore still in the semi-linear regime, even at
subgalactic scales. Using this data it has for instance been
possible to set constraints on the mass of a warm dark matter
particle which makes up all the dark matter \cite{Viel:2005qj}.

\subsection{Telling fermions from bosons}

There is a fundamental difference between hot dark matter of
fermionic and of bosonic nature. First of all, the number and
energy densities are different. For equal values of $\Omega h^2$
this leads to different particle masses and therefore also
different free-streaming behaviour. The differences are at the few
percent level, and although not visible with present data, should
be clearly visible in the future \cite{Hannestad:2005bt}. The
difference between the matter power spectra of two different
models, both with $\Omega_{\rm HDM} = 0.02$, can be seen in
Fig.~2. Even more interesting, in the central parts of dark matter
halos, the density of a bosonic hot dark matter component can be
several times higher than than of a fermionic component with the
same mass, purely because of quantum statistics
\cite{Hannestad:2005bt}. The reason is that the distribution
function, $f=1/(e^{E/T}+1)$, for a non-degenerate fermion in
thermal equilibrium has a maximum at $p=0$ where $f = 1/2$. This
bound also applies to the species after decoupling, and provides
an upper bound on the physical density of such particles in dark
matter halos. This is known as the Tremaine-Gunn bound
\cite{Tremaine:1979we,Kull:1996nx,Madsen:1990pe,Madsen:1991mz}.
Because there is no such limit for non-degenerate bosons, their
density in dark matter halos can be many times higher than that of
fermions. Unfortunately the effect is most pronounced in the
central parts of dark matter halos where the density is dominated
by cold dark matter and baryons, and therefore it might not be
observable \cite{Hannestad:2005bt}.

\begin{figure}[h]
\begin{center}
\epsfysize=5.7truecm\epsfbox{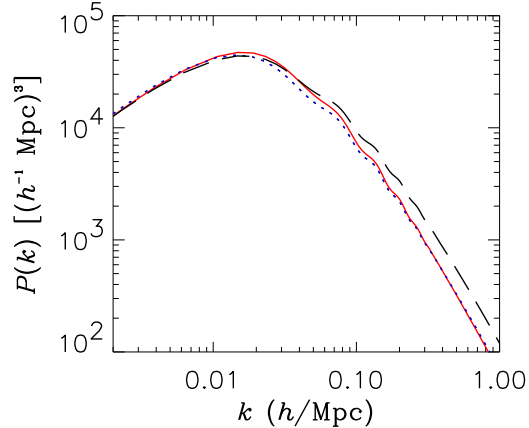} \vspace{0.5truecm}
\end{center}
\caption{Linear power spectra for two different $\Lambda$HCDM
models. The blue (dotted) line shows a model with three massless
neutrinos and one massive Majorana fermion, contributing $\Omega =
0.02$. The red (solid) line shows the same, but with a massive
scalar instead. The black (dashed) line is the standard
$\Lambda$CDM model with no HDM. Note that these spectra have been
normalised to have the same amplitude on large scales. [From
\protect\cite{Hannestad:2005bt}].} \label{fig1}
\end{figure}

%%%%%%%%%%%%%%%%%%%%%%%%%%%%%%%%%%%%%%%%%%%%%%%%%%%%%%%%%%%%%%%%%
\section{Dark energy}
%%%%%%%%%%%%%%%%%%%%%%%%%%%%%%%%%%%%%%%%%%%%%%%%%%%%%%%%%%%%%%%%%

From the present supernova data alone, the universe is known to
accelerate. In terms of the deceleration parameter $q_0$, the
bound is
\begin{equation}
q_0 = - \frac{\ddot a {a}}{\dot a^2} < -0.3
\end{equation}
at 99\% C.L. \cite{Riess:2004}. Such a behaviour can be explained
by the presence of a component of the energy density with strongly
negative pressure, which can be seen from the acceleration
equation
\begin{equation}
\frac{\ddot a}{a} = - \frac{4 \pi G \sum_i (\rho_i + 3 P_i)}{3}.
\end{equation}
The cosmological constant is the simplest (from an observational
point of view) version of dark energy, with $w \equiv P/\rho =-1$.
However, there are many other possible models which produce cosmic
acceleration.

However, since the cosmological constant has a value completely
different from theoretical expectations one is naturally led to
consider other explanations for the dark energy.

\subsection{The equation of state}

If the dark energy is a fluid, perfect or non-perfect, it can be
described by an equation of state $w$ which in principle is
constrainable from observations. Secondly, this dark energy fluid
must have an effective speed of sound $c_s$ which in some cases
can be important.

A light scalar field rolling in a very flat potential would for
instance have a strongly negative equation of state, and would in
the limit of a completely flat potential lead to $w=-1$
\cite{Wetterich:1987fm,Peebles:1987ek,Ratra:1987rm}. Such models
are generically known as quintessence models. The scalar field is
usually assumed to be minimally coupled to matter, but very
interesting effects can occur if this assumption is relaxed (see
for instance \cite{Mota:2004pa}).

In general such models would also require fine tuning in order to
achieve $\Omega_X \sim \Omega_m$, where $\Omega_X$ and $\Omega_m$
are the dark energy and matter densities at present. However, by
coupling quintessence to matter and radiation it is possible to
achieve a tracking behavior of the scalar field so that $\Omega_X
\sim \Omega_m$ comes out naturally of the evolution equation
for the scalar field \cite{Zlatev:1998tr,Wang:1999fa,Steinhardt:1999nw,Perrotta:1999am,%
Amendola:1999er,Barreiro:1999zs,Bertolami:1999dp,Baccigalupi:2001aa,%
Caldwell:2003vp}.

Many other possibilities have been considered, like $k$-essence,
which is essentially a scalar field with a non-standard kinetic
term
\cite{Armendariz-Picon:1999rj,Chiba:1999ka,Armendariz-Picon:2000ah,%
Chimento:2003ta,Gonzalez-Diaz:2003rf,Scherrer:2004au,Aguirregabiria:2004te}.
It is also possible, although not without problems, to construct
models which have $w<-1$, the so-called phantom energy models
\cite{Caldwell:1999ew,Schulz:2001yx,Carroll:2003st,Gibbons:2003yj,%
Caldwell:2003vq,Nojiri:2003vn,Singh:2003vx,Dabrowski:2003jm,Hao:2003th,%
Stefancic:2003rc,Cline:2003gs,Brown:2004cs,Onemli:2002hr,Onemli:2004mb,%
Vikman:2004dc,polarski}.

From an observational perspective there are numerous studies in
which the effective equation of state of the dark energy has been
constrained.

The simplest parametrization is $w=$ constant, for which
constraints based on observational data have been calculated many
times
\cite{Corasaniti:2001mf,Bean:2001xy,Hannestad:2002ur,Melchiorri:2002ux}.
The bound on the equation of state, $w$, assuming that it is
constant is roughly (see \cite{wang04,hm,upadhye,Seljak:2004xh})
\begin{equation}
-1.2 \leq w \leq -0.8
\end{equation}
at 95\% C.L. Very interestingly, however, there is a very strong
degeneracy between measurements of $w$ and the neutrino mass $\sum
m_\nu$. When the neutrino mass is included in fits of $w$ the
lower bound becomes much weaker and the allowed range is
\begin{equation}
-2.0 \leq w \leq -0.8
\end{equation}
at 95\% C.L. \cite{sth2}. The result of a likelihood analysis
taking both parameters to be free can be seen in Fig.~3.

\begin{figure}[h]
\begin{center}
\epsfysize=5.7truecm\epsfbox{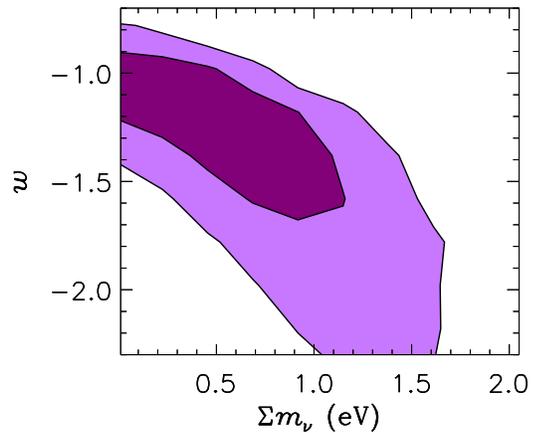} \vspace{0.5truecm}
\end{center}
\caption{The 68\% (dark) and 95\% (light) likelihood contours for
$m_\nu$ and $w$ for WMAP, SDSS, and SNI-a data. [From
\protect\cite{sth2}]} \label{newfig}
\end{figure}

Even though a constant equation of state is the simplest
possibility, as the precision of observational data is increasing
is it becoming feasible to search for time variation in $w$.

At present there is no indication that $w$ is varying. Even though
the present Type Ia supernova data seem to favour a rapid
evolution of $w$, this indication vanishes if all available
cosmological data is analysed
\cite{wang04,hm,upadhye,Seljak:2004xh} (for other discussions of a
time-varying $w$, see for instance
\cite{Corasaniti:2004,Gong:2004a,Gong:2004b,Nesseris:2004wj,%
Feng:2004ad,Alam:2004a,Alam:2004b,Corasaniti:2002,Jassal:2004ej,%
Choudhury:2003tj,Huterer:2004,Daly:2003,Wang:2004a,Wang:2004b,%
Wang:2004c,Jonsson:2004,Weller:2003hw,Linder:2003,pad1,pad2,pad3}.

\subsection{The sound speed of dark energy}

In general the dark energy speed of sound is given by
\begin{equation}
c_s^2 = \frac{\delta P}{\delta \rho},
\end{equation}
if it can be described as a fluid. The perturbation equations
depend on the speed of sound in all components, including dark
energy, and therefore $c_s^2$ can in principle be measured
\cite{Hu:1998tk,Erickson:2001bq,Bean:2003fb,Hannestad:2005ak}.

For a generic component with constant $w$, the density scales as
$a^{-3(1+w)}$, where $a$ is the scale factor. Therefore, the ratio
of the energy density to that in CDM is given by $\rho/\rho_{\rm
CDM} \propto a^{-3w}$. If $w$ is close to zero this means that
dark energy can be important at early times and affect linear
structure formation. If, on the other hand, $w$ is very negative,
dark energy will be unimportant during structure formation. This
also means that since $w \leq - 0.8$ there is effectively no
present constraint on the dark energy equation of state. In Fig.~4
we show current constraints in $w$ and $c_s^2$.

\begin{figure}[h]
\begin{center}
\epsfysize=5.7truecm\epsfbox{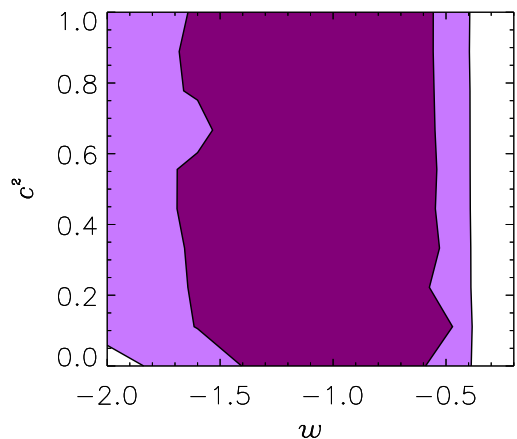} \vspace{0.5truecm}
\end{center}
\caption{The 68\% (dark) and 95\% (light) likelihood contours for
$w$ and $c_s^2$ for WMAP, SDSS, and SNI-a data. [From
\protect\cite{Hannestad:2005ak}].} \label{phantom3}
\end{figure}

\subsection{Dark energy or modified gravity?}

A potentially very interesting possibility is that what we
perceive as dark energy is in fact a modification of gravity on
very large scales. General relativity has been tested to work in
the weak field regime up to supergalactic scales. However, it is
possible that at scales close to the Hubble horizon there might be
modifications.

One possible scenario is that there are extra spatial dimensions
into which gravity can propagate. For instance in the
Dvali-Gabadadze-Porrati model \cite{dgp}, the standard model is
confined to a 3+1 dimensional brane in a 4+1 dimensional bulk
where gravity can propagate. On small scales, gravity can be made
to look effectively four dimensional by an appropriate tuning of
the model parameters, whereas on large scales gravity becomes
weaker. This leads to an effect very similar to that of dark
energy. Based on this idea, other authors have taken a more
observational approach, adding extra terms to the Friedmann
equation \cite{Dvali:2003rk,Elgaroy:2004ne,Ishak:2005zs}.

In this case the dark energy has no meaningful speed of sound
since it is a change in gravity. However, exactly since it affects
gravity it also affects the way in which structure grows in the
universe. In \cite{Hannestad:2004ts} it was found that, unless the
cross-over scale has very specific and fine tuned values, models
with modified large scale gravity are almost impossible to
reconcile with present observations.

\section{Future observations}

\subsection{Cosmic microwave background}

In the coming years, the present CMB experiments will be
superseded by the Planck Surveyor satellite \cite{planck}, due to
be launched in 2007. It will carry instrumentation similar to that
on the latest BOOMERANG flight, but will carry out observations
from space, and for several years. The expectation is that the
project will measure the CMB spectrum precisely up to $l \sim
2500$, being essentially limited only by foreground in this range.
This experiment will be particularly important for the study of
inflation because it will be able to measure the primordial
spectrum of fluctuations extremely precisely.

On a longer timescale there will be dedicated experiments
measuring small scales, such as the Atacama Cosmology Telescope
\cite{Kosowsky:2004sw}. Small scale observations will be
instrumental in understanding non-linear effects on the CMB,
arising from sources such as the Sunyaev-Zeldovich effect and weak
gravitational lensing.

\subsection{Type Ia supernovae}

There are several ongoing programs dedicated to measuring high
redshift supernovae. For instance the Supernova Legacy Survey is
currently being carried out at the CFHT \cite{Pain:2004zc}.
ESSENCE \cite{Matheson:2004dj} is another project dedicated to
improving the current measurement of $w$. The future Dark Energy
Survey \cite{Flaugher:2004vg} is expected to find about 2000 Type
Ia supernovae, and the Supernova Acceleration Probe (SNAP)
satellite mission (one of the contenders for the NASA Dark Energy
Probe program) will find several thousand supernovae out to
redshifts of order 2 \cite{SNAP}.

\subsection{Weak lensing}

Perhaps the most interesting future probe of cosmology is weak
gravitational lensing on large scales. The shape of distant
galaxies will be distorted by the matter distribution along the
line of sight, and this effect allows for a direct probe of the
large scale distribution of the gravitational potential (see for
instance \cite{Bartelmann:1999yn} for a review). Just as for the
CMB the data can be converted into an angular power spectrum, in
this case of the lensing convergence
\cite{Bartelmann:1999yn,Kaiser:1991qi,Kaiser:1996tp,Jain:1996st}.
Several upcoming surveys aim at measuring this spectrum on a large
scale. The first to become operational is the Pan-STARRS
\cite{panstarrs} project which will have first light in 2006. In
the more distant future, the Large Synoptic Survey Telescope
\cite{lsst} will provide an even more detailed measurement of
lensing distortions across large fractions of the sky.

\subsection{The impact on cosmological parameters}

Many of the cosmological parameters will be measured much more
precisely with future data. For the standard cosmological
parameters, a detailed discussion and analysis can be found in
\cite{Ishak:2003zw}. As an example, the bound on the physical
matter density could be improved from the present $\pm 0.012$ to
$\pm 0.0022$, at least an improvement by a factor 5.

With regards to hot dark matter, the neutrino mass could be
constrainable to a precision of $\sigma(\sum m_\nu) \sim 0.1$ eV
or better \cite{hierarchy,future,hu05,hu052,hu053,sth2}, perhaps
allowing for a positive detection of a non-zero mass.

The equation of state of the dark energy could be measurable to a
precision of about 5\%, depending on whether it varies with time
\cite{upadhye}.

\section{Discussion}

We are currently in the middle of an immensely exciting period for
cosmology. We now have estimates of most basic cosmological
parameters at the percent level, something which was almost
unthinkable a decade ago. Cosmology is now at the stage where it
can contribute significant new information of relevance to
particle physics. One notable example is the density of cold dark
matter, which is relevant for SUSY parameter space exploration.
Another is the bound on the mass of light neutrinos which is
presently significantly stronger than the corresponding laboratory
bound.

The precision with which most of the cosmological parameters can
be measured is set to increase by a factor of 5-10 over the next
ten years, given a whole range of new experiments. For the
foreseeable future, cosmology will be an extremely interesting
field, and its relevance to particle physics is set to increase
with time.

\end{document}